\documentclass[prb,aps,twocolumn]{revtex4}
\input epsf
\input pstricks

\usepackage{multirow,amsmath,graphicx,times}

\begin{document}

\title{A minimal representation of the self-assembly of virus capsids}
\author{J.M. Gomez Llorente} 
\email{jmgomez@ull.es}
\author{J. Hern\'andez-Rojas}
\email{jhrojas@ull.es}
\author{J. Bret\'on}
\email{jbreton@ull.es}
\affiliation{ Departamento de F\'{\i}sica Fundamental II and IUdEA, Universidad de
  La Laguna, 38205, La Laguna, Tenerife, Spain}

\begin{abstract}
Viruses are biological nanosystems with a capsid of protein-made capsomer
units that encloses and protects the genetic material responsible
for their replication. Here we show how the
geometrical constraints of the capsomer-capsomer
interaction in icosahedral capsids fix the form of
the shortest and universal
truncated multipolar expansion of the two-body 
interaction between capsomers.  
The structures of many
of the icosahedral and related virus capsids are located as single
lowest energy states of this potential energy surface. Our approach
unveils relevant features of the natural design of the capsids and
can be of interest in fields of nanoscience and nanotechnology 
where similar hollow convex structures are relevant.
\end{abstract}

\maketitle
\section{Introduction}
Viruses are among the simplest biological nano-systems, yet complex
already at the molecular scale. All viruses have a capsid, i.e. a
viral protein shell, that encloses the genetic material (RNA or DNA).
It is formed by subunits called capsomers, which are composed of a
variable number of proteins (from one to six). Despite their diversity,
the coat proteins of many viruses spontaneously self-assemble in vivo
and in vitro into the final viral capsid architecture.

About half of all viruses found so far have capsids with icosahedral
symmetry ($I$ symmetry group) \cite{Zlotnik2005}. In many of these cases, there are 12
capsomers, called pentons, surrounded by 5 neighbour capsomers (i.e.
with coordination number $n=5$); the other capsomers are called hexons
and they have 6 neighbour capsomers ($n=6$). The simple geometrical
construction model introduced by Caspar and Klug (CK) \cite{Caspar1962}
to explain the architecture of icosahedral viruses is a milestone
in modern virology. Generalizations of the CK rules properly account
for the geometry of some exceptional icosahedral capsids \cite{Twarock2004,Keef2006,Lorman2007,Lorman2008}
and other elongated virus capsids that share coordination numbers
with the icosahedral ones \cite{Moody1965,Moody1999,Luque2010}. 
Recent work \cite{Mannige2008,Mannige2009,Mannige2009b} provides 
important further development about the effect of the geometrical 
and topological constraints on the capsid structure.

The knowledge of the capsid interaction potential energy surface (PES)
is a determining step in order to provide a correct description not
only of the self-assembly process but also of the physical properties
of the capsids. Apart from the computationally involved and physically
hardly rewarding full-atom approach \cite{Freddolino2006}, the goal
of most of the work done so far on this matter is a good coarse-grained
model of such PES. The simplest models represent the capsomers as
isotropic bodies, but they require additional geometrical constraints
such as a template of the virus capsid \cite{Bruinsma2003,Zandi2004,Zandi2005,Chen2007a,Chen2007b,Luque2010b}.
In other more complex models each capsomer is represented as a discrete
set of either isotropic \cite{Rapaport2004,Wales2005,Nguyen2006,Chen2007a,Nguyen2008,Fejer2009}
or anisotropic interaction centres \cite{Hagan2006,Douglas2006,Elrad2008},
or as a continuous body of interaction points plus some extra discrete
centres \cite{Fejer2010}.

Many complex systems often organize themselves following rather simple
rules that emerge out of such complexity. The knowledge gathered about
icosahedral virus capsids points towards this kind of simple organization
in these systems. In the present work further 
evidence of this behaviour is revealed by finding the simplest form of a 
two-body interaction, when this is written as an expansion in the complete 
basis set of the angular terms of the standard multipolar expansions \cite{Biedenharn}, 
that incorporates the most relevant geometrical features of the real 
capsomer-capsomer interaction. This form will turn out to be 
universal and will unveil
the important physical factors that determine
the energetic stability of the capsid and its relationship with its
geometrical structure. These factors must somehow
be encoded in the virus genetic material and can become the target
of biological evolution and of drug design.

\section{The interaction model}
Our minimal model represents the capsomer $i$ as an unitary vector
$\boldsymbol{v}_{i}$ at the position vector $\boldsymbol{r}_{i}$
in 3D space. The capsomer plane is supposed to be perpendicular to
$\boldsymbol{v}_{i}$. Since, excluding pentons, the coordination
number for the capsomers in icosahedral capsids is $n=6$ and this
is the natural coordination for 2D isotropic interacting bodies, we
will assume capsomers that are isotropic against rotations around
$\boldsymbol{v}_{i}$. 
The orientational dependence of the interaction between
two capsomers $(i,\, j)$ with relative position vector $\boldsymbol{r}_{ij}=\boldsymbol{r}_{j}-\boldsymbol{r}_{i}$
along $\boldsymbol{n}_{ij}=\boldsymbol{r}_{ij}/r_{ij}$ will be
expanded as a linear combination of the simplest multipolar terms.
Namely, apart from the isotropic term (monopole-monopole), we will
require the monopole-dipole terms $\boldsymbol{v}_{i}\mathord{\cdot}\boldsymbol{n}_{ij}$
and $\boldsymbol{v}_{j}\mathord{\cdot}\boldsymbol{n}_{ij}$, the two dipole-dipole
terms $\boldsymbol{v}_{i}\mathord{\cdot}\boldsymbol{v}_{j}$ and $(\boldsymbol{v}_{i}\mathord{\cdot}\boldsymbol{n}_{ij})(\boldsymbol{v}_{j}\mathord{\cdot}\boldsymbol{n}_{ij})$,
and the monopole-quadrupole contributions $(\boldsymbol{v}_{i}\mathord{\cdot}\boldsymbol{n}_{ij})^{2}$
and $(\boldsymbol{v}_{j}\mathord{\cdot}\boldsymbol{n}_{ij})^{2}$. If the $z$-axis
of a body-fixed reference frame is chosen along $\boldsymbol{n}_{ij}$,
then $\boldsymbol{v}_{i}\mathord{\cdot}\boldsymbol{n}_{ij}=\cos\theta_{i}$,
$\boldsymbol{v}_{j}\mathord{\cdot}\boldsymbol{n}_{ij}=\cos\theta_{j}$, and
$\boldsymbol{v}_{i}\mathord{\cdot}\boldsymbol{v}_{j}=\sin\theta_{i}\sin\theta_{j}\cos(\phi_{i}-\phi_{j})+\cos\theta_{i}\cos\theta_{j}$,
where $(\theta_{i},\,\phi_{i})$ and $(\theta_{j},\,\phi_{j})$ are
the polar coordinates of $\boldsymbol{v}_{i}$ and $\boldsymbol{v}_{j}$,
respectively, in the body-fixed frame.

If the contact between two of our ideal capsomers occurs at the boundary 
of the capsomer planes, its equilibrium geometry will have a two-fold symmetry 
axis bisecting the dihedral angle between the planes. Therefore one must have 
$\phi_{i}-\phi_{j}=0$ and $\theta_{i}+\theta_{j}=\pi$. One can readily prove that 
the the form  $-\boldsymbol{v}_{i}\mathord{\cdot}
\boldsymbol{v}_{j}+\beta\bigl(\boldsymbol{v}_{i}\mathord{\cdot}\boldsymbol{n}_
{ij}\bigr)\bigl(\boldsymbol{v}_{j}\mathord{\cdot}\boldsymbol{n}_{ij}\bigr)$
(rotationally invariant around the intercapsomer axis) with $\beta=2$
provides the unique and universal dipole-dipole interaction
compatible with the 
geometrical constraints (the term is quite different from 
the electrostatic one which changes sign and has $\beta=3$). 
This ansatz does not depend on the dihedral 
angle $\Theta=\theta_{i}-\theta_{j}$.

It is well known that the value of the dihedral angle $\Theta$ is intimately connected to protein 
structure and strongly affects the capsid geometry. 
One can prove that the simplest  
combination of multipolar terms required to fix $\Theta$
must include 
monopole-dipole and monopole-quadrupole terms in the unique form 
$(\boldsymbol{v}_{i}\mathord{\cdot}\boldsymbol{n}_{ij})^{2}+(\boldsymbol{v}_{j}\mathord{\cdot}\boldsymbol{n}_{ij})^{2}+2\cos(\pi/2-\Theta/2)\bigl(\boldsymbol{v}_{i}\mathord{\cdot}\boldsymbol{n}_{ij}-\boldsymbol{v}_{j}\mathord{\cdot}\boldsymbol{n}_{ij}\bigr)$. 

Remarkably, as will be shown later, no other angular terms will be generally required. 
A very convenient form of the capsomer-capsomer interaction
that includes the isotropic and the previously fixed angular terms reads 
\begin{equation}
\begin{split} & V_{ij}=4\varepsilon_{ij}\Bigl\{\left(\frac{\sigma_{ij}}{r_{ij}}\right)^{2m}-\left(\frac{\sigma_{ij}}{r_{ij}}\right)^{m}\\
\times & \Bigl[1+\alpha_{ij}\bigl[\boldsymbol{v}_{i}\mathord{\cdot}\boldsymbol{v}_{j}-\beta_{ij}\bigl(\boldsymbol{v}_{i}\mathord{\cdot}\boldsymbol{n}_{ij}\bigr)\bigl(\boldsymbol{v}_{j}\mathord{\cdot}\boldsymbol{n}_{ij}\bigr)-\zeta_{ij}\bigr]\\- &\frac{1}{2}\gamma_{ij}\bigl[\bigl(\boldsymbol{v}_{i}\mathord{\cdot}\boldsymbol{n}_{ij}+\eta_{ij}\bigr)^{2}+\bigl(\boldsymbol{v}_{j}\mathord{\cdot}\boldsymbol{n}_{ij}-\eta_{ij}\bigr)^{2}\bigr]\Bigr]\Bigr\},
\end{split}
\label{eq:1}
\end{equation}
where $m$ is an integer power and all greek letters are the other
potential parameters. We know that $\beta_{ij}$ is a critical parameter and the 
calculations show that its value can not deviate too much from its theoretical value of 2. 
Somewhat less critical is $\eta_{ij}=\cos(\pi/2-\Theta/2)$, which fixes the dihedral 
angle $\Theta$ when $\gamma_{ij}\neq0$. Besides $\alpha_{ij},\,\gamma_{ij},\,\zeta_{ij}\geq0$ 
is required, the results being quite robust in wide intervals of their values. We have set $\alpha_{ij}=1$. $\zeta_{ij}$ 
is a positive shift parameter that affects the intercapsomer 
equilibrium distance; for $\beta_{ij}=2$ its most convenient value is $\zeta_{ij}=1$ and,
in this case, the binding energy of
the capsomer dimer is
the same independent of the values chosen for the parameters
$\alpha_{ij}$, $\gamma_{ij}$ and $\eta_{ij}$, and its equilibrium
distance is always $r_{ij}^{\mathrm{e}}=2^{\frac{1}{m}}\sigma_{ij}$.
The value of the power $m$ in Eq. \ref{eq:1} is not of critical
importance either \cite{Supp}. We have chosen for most calculations $m=12$.
The total interaction energy for a set of $N$
capsomers will be evaluated as the sum of all pair interactions, namely
$V=\frac{1}{2}\sum_{i\neq j}^{N}V_{ij}$.

We have performed calculations with one type (hexons) and two types
(pentons and hexons) of capsomers. For the hexon-hexon
potential we have set $\varepsilon_{ij}=\varepsilon_{\mathrm{hh}}=1$
as the energy unit and $\sigma_{ij}=\sigma_{\mathrm{hh}}=1$ as the
length unit. 
In general, we have set $\gamma_{ij}=\gamma_{\mathrm{hh}}=0$, thus the
hexon-hexon dihedral angle is left free so that it can adapt to non
equivalent geometrical environments without frustration. 
For capsids with only hexons, a non-zero 
value of $\gamma_{ij}$ may be required to favour the formation of a
given capsid structure, and in such cases 
we have chosen $\gamma_{\mathrm{hh}}=5$. 
The penton-penton potential parameters are not critical; our 
choice for them was $\varepsilon_{ij}=\varepsilon_{\mathrm{pp}}=0.1$,
$\sigma_{ij}=\sigma_{\mathrm{pp}}=1$ and $\gamma_{ij}=\gamma_{\mathrm{pp}}=0$.

When pentons are present, 
each penton should have five hexon neighbours. These local structures
are critical in the self-assembly of hollow capsids and we have fixed
the parameters of the hexon-penton pair potential to favour their formation;
namely we have chosen $\varepsilon_{ij}=\varepsilon_{\mathrm{hp}}=1.5$
and $\sigma_{ij}=\sigma_{\mathrm{hp}}=2^{\frac{m-1}{m}}R\sin\vartheta_{\mathrm{hp}}$,
with $\vartheta_{\mathrm{hp}}=\frac{1}{2}\arcsin[\sigma_{\mathrm{hh}}/(2^{\frac{m-1}{m}}R\sin\frac{\pi}{5})]$;
with the latter value for $\sigma_{\mathrm{hp}}$ the pentagonal pyramid
formed by a central penton surrounded by five hexons located at the
vertices of a regular hexagon whose edge lengths are the corresponding 
capsomer-capsomer equilibrium distances 
will have all its vertices lying on the surface of a sphere whose
radius is $R$.
Although not necessary in most of the capsid structures found, 
surface curvature can be further fixed through the hexon-penton 
interaction by choosing a non vanishing $\gamma_{ij}$. 
We have generally chosen $\gamma_{ij}=\gamma_{\mathrm{hp}}=5$ 
and $\eta_{ij}=\eta_{\mathrm{hp}}=2^{\frac{1-m}{m}}\sigma_{\mathrm{hp}}/R$;
this last choice induces a curvature radius $R$ for an hexon-penton
pair at its equilibrium distance. The same value was also used
for $\eta_{\mathrm{hh}}$ in capsids made only of hexons. 

The only free parameter in our potential
model is therefore $R$. Obviously $R$ should depend on the total
number of capsomers, $N$, in the capsid. We have varied it with the
simple relation $R=2^{1/m}\sigma_{\mathrm{hh}}\left[\frac{\sqrt{3}(N-2)}{8\pi}\right]^{1/2}$,
which is obtained from the surface of the unfolded capsid icosahedron
on the 2D hexagonal lattice.

\section{Results and discussion}
\begin{figure}[h]
\includegraphics{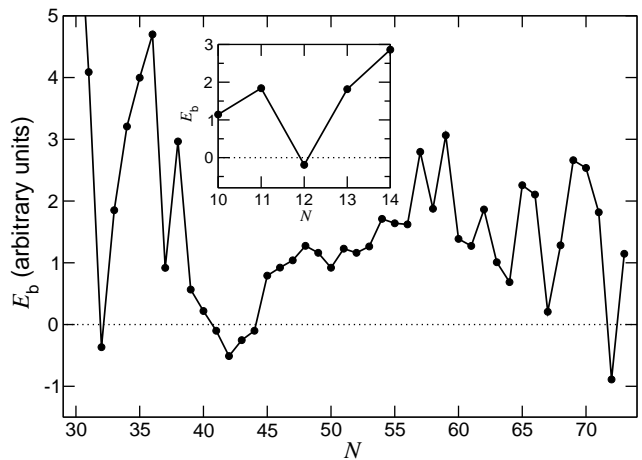}\caption{\label{fig:1}Capsid binding energies $E_{\mathrm{b}}$ as a function
of the capsomer number $N$. For the main curve the number of pentons
is fixed to 12. For the inset figure all capsomers are of the same type (hexons in our notation).}
\end{figure}
We will name the different capsids, when possible, with their corresponding
triangulation numbers $T$ \cite{Caspar1962,Caspar1980}, $Q_{\mathrm{5F}}$
\cite{Moody1965,Moody1999,Luque2010}, $Q_{\mathrm{3F}}$, $Q_{\mathrm{2F}}$
\cite{Luque2010}. Basin-hopping global optimization \cite{Wales,Li,Wales2}
was used to find the lowest energy structures.
We have defined conveniently shifted binding energies 
$E_{\mathrm{b}}=V+(3N-6)\varepsilon_{\mathrm{hh}}+5N_{\mathrm{p}}(\varepsilon_{\mathrm{hp}}-\varepsilon_{\mathrm{hh}})$, $N_{\mathrm{p}}$ being the number of pentons.
With this choice a capsid with optimal packing and contact interactions
would have $E_{\mathrm{b}}=0$.
Values of $E_{\mathrm{b}}$ slightly lower than zero are possible
due to the (small) long-range tail of our interaction. $E_{\mathrm{b}}$ is also a natural and convenient measure of frustration 
for the class of viral
capsids considered in this work; namely, capsids with
$E_{\mathrm{b}}\lesssim0.5$ present low frustration structures,
i.e. structures with optimal coordination for each capsomer and
pair interaction energies $V_{ij}$ between neighbour capsomers close
to their minimum values.

The calculations performed using only one type of capsomers
(i.e. hexons) and the simplest form of the interaction with free 
dihedral angle (i.e. $\gamma_{\mathrm{hh}}=0$) provide
hollow lowest-energy structures \cite{Supp}.
Particularly stable capsids are those with icosahedral symmetry
corresponding to the allowed triangulation numbers $T$. 
There are indeed real capsids made of only one type of capsomers \cite{Supp}
and some of our forms have been reproduced 
with other theoretical models \cite{Zandi2004,Chen2007a,Chen2007b,Fejer2010,Johnston2010}. 
However in this case competing structures appear close in energy
to the global minima, this behaviour being a consequence of the
relatively large frustration of those minima, with energies
$E_{\mathrm{b}}\gtrsim10$ that increase with $N$.
Such competition can be reduced significantly for icosahedral 
capsids by choosing $\gamma_{\mathrm{hh}}\neq0$ and
an adequate dihedral angle. Nevertheless, this procedure never 
reduces frustration. The only way to do so is by replacing the hexons 
with coordination $n=5$ by pentons with the right structure. Thus 12 
pentons are required in agreement with Nature's choice for most of the 
icosahedral an related virus capsids. 
Henceforth we will focus our attention on these capsids 
made of hexons and 12 pentons. 

\begin{figure}
\includegraphics{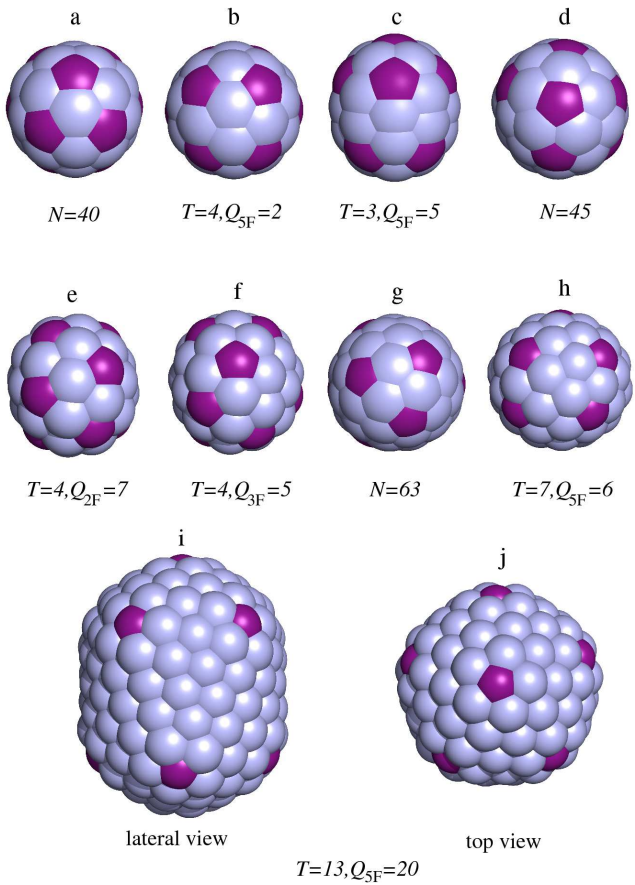}\caption{\label{fig:3}The structure of non-icosahedral capsids
found as global energy minima in our PES. Purple capsomers are pentons while light-blue ones are hexons.}
\end{figure}

Figure \ref{fig:1} presents the global minimum energies 
($E_{\mathrm{b}}$) as a function of the capsomer number $N$.
A relevant property
of our model is that if the curvature parameter $R$ is included
in the global optimization procedure, its stabilizing effect is significantly
more important for the icosahedral structures. 
\begin{figure}[h]
\includegraphics{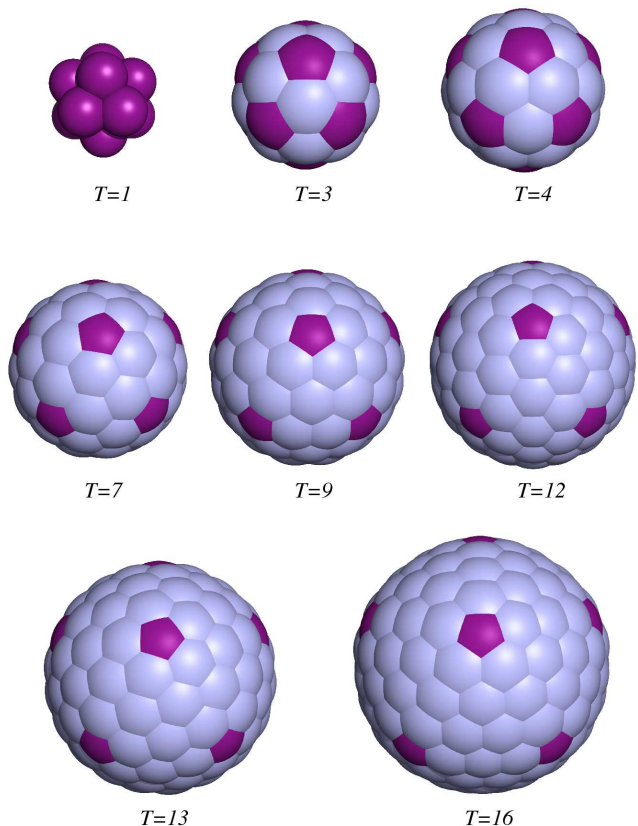}\caption{\label{fig:2} 
The structure of icosahedral capsids
found as global energy minima in our PES. Capsomers with
coordination number $n=5$ are purple. With our parameter choice, 
these are hexons for
the $T=1$ capsid and pentons for all other capsids. Light-blue capsomers are all hexons.}
\end{figure}
Capsids corresponding to the allowed triangulation numbers $T$ up
to $T=16$ (some of these outside the range in Figure \ref{fig:1}),
have the icosahedral structures as the lowest energy and optimal packing
structures, having binding energies closer to or smaller than zero
and also lower than those of their neighbours in capsomer number. Illustrations
of these icosahedral capsids are shown in Figure \ref{fig:2}. In
the $T$ range studied, we found a relevant exception to the previous
rule for $N=42$, which would correspond to a capsid with $T=4$.
The lowest energy structure found in this case corresponds to a non icosahedral spheroidal geometry
given by the two triangulation numbers $T=4$, $Q_{\mathrm{5F}}=2$,
and having the symmetry of the $D_{5}$ point group (see Figure \ref{fig:3}b). The corresponding
$T=4$ capsid with icosahedral geometry has an energy only $0.01$
units above the global energy minimum. Both capsids have practically
the same optimal $R$, and therefore no discrimination between them
is possible by changing the curvature. This result points to one of
two possible consequences: either the virus capsid could adopt both
structures without functional implications, or additional terms would 
be required in our interaction model to include geometrical constraints
that would eliminate one of
the competing geometries without adding significant frustration to
the other. The higher symmetry of the icosahedral capsid makes much easier 
the discrimination in favour of this structure. Namely, we have been
able to remove the competition of the $D_{5}$ capsid by adding some
terms to the penton-hexon and hexon-hexon interactions that incorporate
the $C_{2}$ symmetry of hexons in the icosahedral capsid (details
will be presented elsewhere). One can easily find that similar non
icosahedral but spheroidal structures are possible for all $T$ numbers
that do not coincide with the corresponding class number $P$ \cite{Supp}.

Another optimal packing structure corresponding to $N=42$ is that
of the $\Phi29$ virus capsid. This has a prolate geometry with $T=3$
and $Q_{\mathrm{5F}}=5$ \cite{Tao1998}. Our model is also able to
provide this structure (Figure \ref{fig:3}c), which can be made the 
lowest energy structure by optimizing its curvature $R$ after 
conveniently increasing the value of the 
$\gamma_{\mathrm{hp}}$ parameter. 

For capsomer numbers $N$ between the values corresponding to the
allowed triangulation numbers $T$, we find capsids with optimal packing
but higher frustration. Examples are included in Figure \ref{fig:3}.
Some of them follow Moody's \cite{Moody1965,Moody1999} (oblate for $N=67$
in Figure \ref{fig:3}h) and Luque and Reguera's \cite{Luque2010}
rules ($N=46$ and $48$ in Figure \ref{fig:3}e and f, respectively),
but others are completely new ($N=40$ with tetrahedral symmetry in
Figure \ref{fig:3}a, $N=63$ with $D_{3}$ symmetry in Figure \ref{fig:3}g,
or the asymmetric $N=45$ in Figure \ref{fig:3}d). Figure \ref{fig:1}
shows particularly stable capsids for $N$ in the neighbourhood of
the $T=4$ capsid (e.g. $N=41,43,44$) that have spheroidal geometries
of the new types with some remaining symmetry (e.g. $D_{3}$, $C_{2}$,
$S_{4}$, respectively).

The larger capsid that we have been able to find is that of the prolate
bacteriophage T4 head. This has a capsomer number $N=167$, with $T=13$
and $Q_{\mathrm{5F}}=20$ \cite{Fokine2004}. Two views of this capsid
are shown in Figure \ref{fig:3}(i and j). After optimizing the curvature
$R$, the energy for this capsid is $E_{\mathrm{b}}=5.97$, which
indicates an important frustration. Similar behaviour for the frustration
is also found in the larger icosahedral capsids $T\geq9$, in agreement
with geometrical considerations \cite{Mannige2009}. By slightly modifying
our potential model \cite{Supp}, we have been able to incorporate
in it the known buckling process \cite{Wikoff2006} and reduce significantly
the frustration. The illustrations for this capsid in Figure \ref{fig:3}
include this process, which is particularly evident in its top view.

In order to explore the available capsid configuration space as a
function of temperature $T^{*}$, we have used parallel tempering 
Monte Carlo sampling \cite{Geyer1991,Earl2005,Neirotti2000} of such space for the $T=3$
capsid \cite{Supp}. First of all, let us point out that since our
model assumes unstructured capsomers, it can not account for processes
known to affect the structure of such units \cite{Ausar2006}. However
this sampling provides relevant information about possible capsid
structures before the disassembly transition. The ideal caloric curves
obtained from this statistical analysis \cite{Supp} show the capsid
disassembly as a first-order-like phase transition with a sharp change
in the internal energy. Moreover, at temperatures below this transition,
we only find vibrationally excited forms of the optimal capsid structure.

\section{Conclusions}
By taking into account geometrical constraints, 
we have derived the shortest and yet quite universal form of
a truncated multipolar expansion for the capsomer-capsomer 
pair interaction in icosahedral capsids.
For capsids made of only one kind of capsomers a simple form
without adjustable parameters provides a great variety of hollow 
structures. The simplicity and universality 
of this form makes it a promising tool in the understanding of the
physical properties of homogeneous hollow structures of relevance 
in many applied fields of nanoscience and nanotechnology 
(membranes, colloidal surfaces, etc.).
The introduction of a different type of capsomers (12 pentons)
eliminates competing structures and reduces frustration significantly.
In this case, the curvature $R$ is the only relevant parameter of the model.
By varying this parameter and the number of capsomers $N$, 
the model is able to provide many optimal packing structures
as global energy minima. Apart from the allowed icosahedral structures,
we find geometries satisfying Moody's \cite{Moody1965,Moody1999} and Luque
and Reguera's rules \cite{Luque2010}, and other completely new structures.
This simple form unveils the following relevant physical features
of the natural design of the viral capsids.

The stabilizing effect of an optimal choice of the curvature is significantly
larger for the icosahedral capsids. Besides, prolate structures (e.g.
$\Phi29$ virus) can be stabilized in this way against competing icosahedral
capsids (e.g. $T=4$).

After curvature optimization (the $T=4$ capsid requiring also additional
interaction terms), self-assembly is dominated
and guided by a single capsid structure.

For the larger capsids ($T\geq9$ or $N>90$) we observe an increase
of frustration even after curvature optimization. Buckling has been
shown to reduce such frustration.

There is still a last conclusion of fundamental relevance. This concerns
the physical reasons behind the simplicity yet general validity of
our interaction model. Of course, the highest possible symmetry of
icosahedral capsids is a first main reason; the second one is the
2D close-packing arrangement of the capsomers, which all except the
pentons have coordination number $n=6$. As a matter of fact this
packing is the easiest one to reproduce by pair interaction models.
These two reasons are related by the fact that of all finite rotation
point groups, $I$ is the one that has more symmetry operations in
correspondence with those of the close-packing 2D hexagonal lattice.

Another important physical reason behind simplicity is in the principle
of low frustration satisfied by many-body systems with a unique lowest
energy structure, as should occur, for obvious reasons, in the case
of viral capsids. In an appropriate expansion of the system potential
interaction energy (as the multipolar one used here),
the principle of low frustration seems to reduce
the number of terms required to reproduce the optimal structure.
Of course, many  
terms will have to be added to our universal form to 
approach a particular real interaction,
but these should hardly affect both the capsid structure and its binding
energy. They will certainly change the statistical weight of the configuration
space around that structure and perhaps eliminate other competing geometries,
as we have shown for the $T=4$ capsid.

\section*{Acknowledgement}
We thank Prof. David Reguera and Prof. David J. Wales for their valuable comments on this work.

\footnotesize{
\providecommand*{\mcitethebibliography}{\thebibliography}
\csname @ifundefined\endcsname{endmcitethebibliography}
{\let\endmcitethebibliography\endthebibliography}{}

}

\newpage

\section{Supplementary Material}

In the Article we use, when possible, the notation introduced by Caspar
and Klug to label the icosahedral capsids. These authors build a geometrical
model of the capsid by folding a 2D hexagonal lattice whose equivalent
positions are given by a pair of non negative integers ($h$, $k$).
Then the allowed structures are those having the triangulation number
$T=h^{2}+k^{2}+hk$, (whose possible values are $T=1,3,4,7,9,12,13,..$).
The number $T$ can be written as $T=Pf^{2}$,
where $f$ is the greatest common divisor of $h$ and $k$, and each
allowed $P$ value corresponds to a capsid class. With this construction
the number of hexons is $10(T-1)$ and one always has 12 pentons.
Some other structures have been denoted with one additional triangulation
number $Q_{\mathrm{5F}}$ (see References 7-9 in the Article), $Q_{\mathrm{3F}}$
or $Q_{\mathrm{2F}}$ (see Ref. 9 in the Article).

An important mathematical result has guided us in the construction
of our capsomer-capsomer interaction model: the anisotropic terms
(those depending on the angular orientation) of the standard multipolar
expansion of the interaction between two charge distributions form
a complete basis set to expand the angular dependence of the interaction
between any two bodies. This basis set is the direct product of two
Wigner matrix basis sets (Ref. 29 in the Article),
one for each of the two bodies.

Notice that in our interaction energy expression $\boldsymbol{v}_{i}$
is not an electric dipole, but a vector giving the capsomer orientation.
Of course, the origin of these interaction terms has to be found in
the inter-residue forces existing between the proteins forming each
capsomer. These forces have not only electrostatic contributions,
but also dispersive, inductive and thermodynamical (through the water
solvent) ones. 

The parameter $m$ in our model (Eq. 1 in the Article) determines 
somehow the potential range relative to the capsomer size. 
Interactions between real viral capsomers are thought to be very short-ranged,
requiring values of this parameter as high as $m\sim50$.
We have chosen for most calculations a lower value ($m=12$) for practical
computational reasons.

Notice that if in the expression giving the interaction energy (Eq.
1 in the Article) $\alpha_{ij}$ and $\gamma_{ij}$ are set to zero and $m=6$, then
$V_{ij}$ takes the form of the Lennard-Jones potential with $\varepsilon_{ij}$
being the depth of the interaction well and $\sigma_{ij}$ the finite
distance below which $V_{ij}>0$.

Our results showed that frustration is significantly important for
the larger capsids, e.g. the bacteriophage T4 head and the icosahedral
capsids with $T\geq9$. We also mentioned that by modifying our interaction
model to allow for the buckling process such frustration could be
reduced. Let us explain now this modification. The faceting induced
by buckling in the capsid affects particularly to the hexons at the
edges of the icosahedron. These have to bend somehow to adapt to the
orientation of the two corresponding faces. These bending degrees
of freedom have been introduced by adding two new unitary vectors
$\{\boldsymbol{v}_{ik},\: k=1,2\}$ for each hexon, being their orientation
the 4 new degrees of freedom (more details will be given elsewhere).
Then, the penton-hexon interaction was left unchanged using the original
$\boldsymbol{v}_{i}$, and the anisotropic part, $V_{ij}^{(\mathrm{a})}$,
of the hexon-hexon interaction (remember that $\gamma_{\mathrm{hh}}=0$)
was changed to

\begin{equation}
\renewcommand{\theequation}{S-\arabic{equation}}%
\begin{split}&V_{ij}^{(\mathrm{a})}=-4\varepsilon_{\mathrm{hh}}\left(\frac{\sigma_{\mathrm{hh}}}{r_{ij}}\right)^{m}\Bigl\{\max_{k,l=1,2}\bigl\{1+\alpha_{\mathrm{hh}}\\
\times&\bigl[\boldsymbol{v}_{ik}\cdot\boldsymbol{v}_{jl}-\beta_{\mathrm{hh}}\bigl(\boldsymbol{v}_{ik}\cdot\boldsymbol{n}_{ij}\bigr)\bigl(\boldsymbol{v}_{jl}\cdot\boldsymbol{n}_{ij}\bigr)-\zeta_{\mathrm{hh}}\bigr]\bigr\}\\
+&\frac{1}{2}\kappa_{\mathrm{hh}}\sum_{k=1}^{2}\bigl[(\boldsymbol{v}_{i}-\boldsymbol{v}_{ik})^{2}+(\boldsymbol{v}_{j}-\boldsymbol{v}_{jk})^{2}\bigr]\Bigr\},
\end{split}
\end{equation}
 where $\max$ stands for the maximum value of all $(k,l)$ pairs.
A penalty energy for this bending has been added with the term proportional
to $\kappa_{\mathrm{hh}}$. The penalty parameter is $\kappa_{\mathrm{hh}}>0$,
and for $\kappa_{\mathrm{hh}}\rightarrow\infty$ this expression reduces
to the form included in Eq. 1 of our Article.

Let us comment now on the physical meaning of the arbitrary energy
unit used in our model. From the knowledge of the interaction between
capsomers at the molecular level (S. Reddy, H. A. Giesing, R. T. Morton,
A. Kumar, C. B. Post, C. L. Brooks, III and J. E. Johnson, {\em Biophys.
J.}, 1998, \textbf{74}, 546-558), an estimate of the actual value of
this energy unit is $\varepsilon=1\sim10$ $k_{\mathrm{B}}T^{*}_{\circ}$
units or even higher ($k_{\mathrm{B}}$ is the Boltzmann constant
and $T^{*}_{\circ}$ the normal temperature). This would imply, for instance,
that any other capsid structure differing from the lowest energy one
in just $\sim0.2$ of our energy units would be statistically irrelevant
at normal temperatures.

Except for capsids in the neighborhood of the $T=1$ ($N=12$) capsid, 
we have presented in the Article results obtained for a fixed 
number (12) of pentons and a changing number of hexons. 
However we have also performed more extensive calculations
using only one type of capsomers. In this case we have used the parameters 
of the hexon-hexon interaction with vanishing and non-vanishing 
$\gamma_{\mathrm{hh}}$ values. In both cases we obtain hollow capsids, 
with a larger diversity of forms in the case $\gamma_{\mathrm{hh}}=0$. 
A sharper energetic discrimination between them requires a non-vanishing 
$\gamma_{\mathrm{hh}}$ and an adequate choice of the curvature parameter $R$. 
In this way we could be able to generate capsids sharing the structure of 
those obtained with the inclusion of the two types of capsomers.
A particular example is the structure of the human papilloma virus capsid, 
which is known to be made of just one type of capsomers. Our model is
able to also reproduce the structure of this particular capsid. Its
theoretical triangulation number is $T=7$ ($N=72$). Besides our structure
shows, as the experimental reconstruction (T. S. Baker, W. W. Newcomb,
N. H. Olson, L. M. Cowsert, C. Olson and J. C. Brown, {\em Biophys.
J.}, 1991, \textbf{60}, 1445-1456), some slight asymmetries.
However, as discussed in the Article, this homogeneneous structure
has a relativeley large frustration, with a bonding energy $E_{\mathrm{b}}\simeq20$. By replacing the 12 hexons with coordination $n=5$ by pentons this energy can be reduced to a value $E_{\mathrm{b}}\simeq-1$.

The capsid likely lowest energy structures provided by our interaction
potential model were located using the basin-hopping scheme (Ref.
32 in the Article), which is also known as the ``Monte Carlo plus energy minimization''
approach of Li and Scheraga (Ref. 33 in the Article). This method is very well
known and has been used successfully in the fields of clusters, biomolecules
and glasses (Ref. 34 in the Article).

Our parallel tempering Monte Carlo sampling calculations make use of standard
methods (References 39-41 in the Article). Figure \ref{fig:1} gives, with a continuous
line, the ideal caloric curve for the $T=3$ icosahedral capsid obtained
with the potential-range parameter $m=12$. The ideal transition temperature
is $k_{\mathrm{B}}T^{*}=0.43$ in this case. We have argued that from
the relative length scales of the capsomer size and the interaction
range values as high as $m\sim50$ would be required. While the particular
choice of this parameter hardly affects the structures of the global
energy minima (we have tested choices in the range $6<m<48$ which
provide rather similar structural results), it has significant effects
on the theoretical disassembly transition. Namely for $m=48$ this
ideal transition temperature is reduced to $k_{\mathrm{B}}T^{*}=0.32$
(dashed line in Figure \ref{fig:1}).
\begin{figure}
\renewcommand{\thefigure}{S-\arabic{figure}}%
\includegraphics{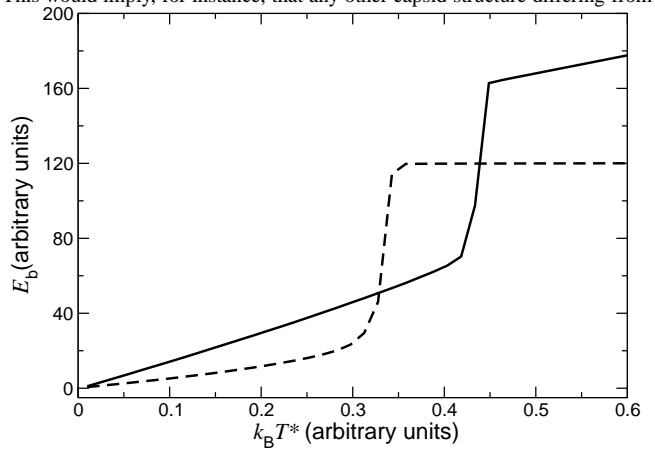}
\caption{\label{fig:1}Caloric curve giving the average internal energy $E_{b}$
as a function of temperature $T^{*}$ ($k_{\mathrm{B}}$ is the Boltzmann
constant) for the icosahedral capsid with triangulation number $T=3$
($N=32$). The continuous line was obtained with a value of the range
parameter $m=12$ (Eq. 1 in the Article), while the dashed line is for $m=48$.}
\end{figure}
 

\end{document}